\documentstyle[epsf,graphicx]{elsart}
\begin{document}
\begin{frontmatter}
\title{ A variational approach to approximate particle number projection
with effective forces. }
\author{ A. Valor\thanksref{kk}}, 
\author{ J.L. Egido and L.M. Robledo }
\address{Departamento de F\'{\i}sica Te\'orica C-XI\\
Universidad Aut\'onoma de Madrid, E--28049 Madrid, Spain}
\thanks[kk]{Present address~: Service de Physique Nucl\'eaire Th\'eorique,
       U.L.B.-C.P.229, B-1050 Brussels, Belgium}
\date{\today}
\maketitle

\begin{abstract}
  Kamlah's second order method for approximate
particle number projection is applied for the first time to 
variational  calculations with effective forces.  High spin states of normal and
 superdeformed nuclei have been calculated  with the finite range 
density dependent Gogny force for several nuclei. Advantages and drawbacks
 of the Kamlah second order method as compared to the Lipkin-Nogami recipe are 
 thoroughly discussed. We find
that the Lipkin Nogami prescription occasionally may fail to find the right energy
minimum in the strong pairing regime and that  Kamlah's second order
approach though providing
better results than the LN one may break down in some limiting situations. 
\end{abstract}
\begin{keyword}
Kamlah-Lipkin-Nogami approximation, Gogny-Interaction, 
Superdeformed-Bands, A=152,164,190.

 21.10.Re, 21.10.Ky, 21.60.Ev, 21.60.Jz, 27.70.+q, 27.80.+w
\end{keyword}
\end{frontmatter}
\section{Introduction}
In recent  years it has become clear that a proper treatment of
pairing correlations is required  to describe many nuclear
properties. In particular, in situations of weak pairing one has to go 
beyond the mean field approximation.  Since exact particle number projection (PNP) is 
very time consuming for non-separable interactions, the approximation  most
widely  used  is the Lipkin-Nogami (LN) 
\cite{LIP.60,NOG.64,GN.66,QUE.90} one, just because of simplicity. 
The LN prescription can be derived in several ways \cite{ZSF.92},  one
transparent derivation \cite{VER.96,VER.99} can be obtained in terms of the Kamlah
 \cite{KAM.68} expansion  to second order. Here one  sees that the
 basic assumptions are~: first, to have a well pair-correlated system 
(i.e. a large particle number fluctuation $\langle (\Delta \hat{N})^2 \rangle$)
  and second, to keep constant
during the energy minimization procedure, the coefficient  ($h_2$) of 
 the particle number fluctuation term of the energy expansion.
 To maintain  $h_2$ constant  amounts to violate
the Ritz variational principle, this 
has advantages and drawbacks.  As we shall see, keeping $h_2$  constant 
amounts to stay in
 the pair-correlated regime, i.e. large $\langle (\Delta \hat{N})^2 \rangle$,
ensuring thereby a good convergence of the Kamlah expansion. 
On the other hand the lack of variational character
may cause to end up in the wrong minimum. 
The restriction of keeping 
 $h_2$ constant can be released and one ends up
with a full variational problem, what we shall call the selfconsistent
 Kamlah  approximation to second order (SCK2).  In principle, a
full variational approach to the problem seems to be the optimal one. 
Under some conditions, however, it may  not be
a good approximation because the variation of $h_2$ may
 lead to  regions of the Hilbert space with small 
 $\langle (\Delta \hat{N})^2 \rangle$ values, 
making the Kamlah expansion to break down.
 
  The validity of both  approaches has  motivated   some interest.
   Zheng, Sprung and Flocard \cite{ZSF.92}, using 
a very simple  model space of two levels  discussed 
the LN and  SCK2 approximation to particle number projection. The force used
in the calculations was a monopole pairing  which 
intensity $G$ could be varied in order to simulate  different pairing
regimes. The energies obtained by means of the two approximate methods
were compared with the exact solutions. As a
result, it was stated that both methods display an excellent agreement
with the exact solution in the regimes of strong and medium 
 pairing correlations. In the regime of weak
correlations, however,  both theories provided energies deeper
than the exact one,  the LN approach remaining  closer, in general, to 
the exact value than the SCK2 approach.
In a comment to this article, Dobaczewski and Nazarewicz \cite{DN.93} 
discussed further on the problem.  In particular, they explained
the strange behavior of the energies in the weak-pairing regime in
terms of a violation of the domain of applicability of a quadratic
approximation in this regime. 
  It is not obvious  that the findings with the exact soluble models will apply
to Hamiltonians with more degrees of freedom. 

 The purpose of this paper is to investigate the LN approach and
the fully variational Kamlah method, SCK2, with effective
forces and large configurations spaces to analyze under which conditions
these approaches are a good approximation to  exact particle number projection.
We shall furthermore analyze the consistency of the Kamlah expansion.
In the numerical application we have solved the  LN and SCK2 equations
 with the finite range density dependent Gogny force \cite{GOG.75}.
We think that the Gogny force is the ideal one to study these effects because,
at variance with the Skyrme and  separable forces,
in addition to the particle-hole interaction
 its finite range  provides also the particle-particle interaction. In the calculations we shall look for situations 
where the plain mean field approach is not expected to provide a good description, as
high spin states and superdeformed states.

\section{Theoretical approximations}

In order to discuss the approximations involved in the LN prescription
and in the SCK2 approach we shall make a short derivation of the theory 
\cite{RS.80,VER.96,VER.99}.
 In the case of large particle number violating HFB wave functions $|\Phi\rangle$, 
i.e. large $\langle (\Delta \hat{N})^2 \rangle$,
with $\Delta\hat{N} = \hat{N} - \langle \Phi | \hat{N}|\Phi \rangle$
and $\langle \hat{N} \rangle \equiv \langle \Phi | \hat{N}|\Phi \rangle $,
 one may use the Kamlah expansion to calculate the PNP energy. 
The projected energy to second order is given by
\begin{equation}
E^{(2)}_{proj}\;=\;\langle\hat{H}\rangle\;\;-\;\;h_{2}\langle(\Delta\hat{N})^{2}\rangle
\;\;+\;\; h_{1}\;(N_0 - \langle\hat{N}\rangle) \;\;+\;\; 
h_{2}\; (N_0 -\langle\hat{N}\rangle)^{2},
\label{energy2}
\end{equation}
and the coefficients $h_1, h_2$ by 
\begin{eqnarray}
  h_{1} & = & \frac{\langle\hat{H}\Delta\hat{N}\rangle - h_{2} \;\langle(\Delta\hat{N})^{3}\rangle}
{\langle(\Delta\hat{N})^{2}\rangle}  \label{h_1}\\
  h_{2} & = & \frac{\langle(\hat{H}-\langle\hat{H}\rangle)(\Delta\hat{N})^{2}\rangle  - 
\langle\hat{H}\Delta\hat{N}\rangle\langle(\Delta\hat{N})^{3}\rangle/\langle(\Delta\hat{N})^{2}\rangle}
{\langle(\Delta\hat{N})^{4}\rangle - \langle(\Delta\hat{N})^{2}\rangle^{2} 
- \langle(\Delta\hat{N})^{3}\rangle^{2}/\langle(\Delta\hat{N})^{2}\rangle}.
\label{h_2}
\end{eqnarray}
The expressions above are valid for one kind of nucleons. In the general case
we have $h_1^p, h_1^n$ and $h_2^p, h_2^n$ for protons and neutrons.
In  a full variation after projection method one  should vary $E^{(2)}_{proj}$.
 As an approximation to this method, in order to avoid to compute
cumbersome expressions like $ \frac{\delta h_2}{\delta \Phi}$ one has used
 the Lipkin-Nogami prescription in which the coefficient $h_2$ is held constant during
the variation.  As a result  the LN variational equation is much simpler, one gets
\begin{equation}
\frac{\delta}{\delta \Phi}\langle \hat{H} \rangle - h_1 \frac{\delta}{\delta \Phi}\langle 
\hat{N}\rangle +
(N_0 - \langle\hat{N}\rangle)\frac{\delta h_1}{\delta \Phi} -2 h_2 (N_0 - 
\langle\hat{N}\rangle)\frac{\delta}{\delta \Phi}\langle\hat{N}\rangle -
 h_2 \frac{\delta}{\delta \Phi} \langle(\Delta\hat{N})^{2}\rangle=0.
\end{equation}
 It is obvious that the solution to this equation is equivalent to solve
\begin{equation}
\frac{\delta}{\delta \Phi}\langle \hat{H} - h_2 (\Delta\hat{N})^{2}\rangle
- \lambda \frac{\delta}{\delta \Phi}\langle \hat{N}\rangle = 0,
\label{cran}
\end{equation}
 with  $\lambda$ determined by  the constraint
\begin{equation}
 \langle \hat{N} \rangle = N_0,
\label{concran}    
\end{equation}
{\it provided} the condition $\lambda =  h_1$ is accomplished.  This can be easily 
checked noticing
that eq.~(\ref{cran}) must hold for any variation $|\delta \Phi \rangle$. In 
particular, we
can choose $|\delta \Phi \rangle = \Delta \hat{N} |\Phi \rangle$,  the
substitution of this specific variation in eq.~(\ref{cran}) provides
\begin{equation}
 \langle \hat{H} \Delta \hat{N} \rangle - \lambda \langle (\Delta \hat{N})^2 \rangle -
 h_2 \langle (\Delta \hat{N})^3 \rangle =0,
\end{equation}
 comparison with eq.~(\ref{h_1})  shows that $ \lambda = h_1 $.
In the Lipkin-Nogami approach though $h_2$ is not varied during the minimization process
it is updated in each iteration of the minimization process.

  If $h_2$ is allowed to vary, the variational principle on $E_{proj}^{(2)}$
  is not anymore 
equivalent to minimize eq.~(\ref{cran}) with the constraint on the particle number,
 because in this case $ \lambda \neq h_1 $. We can, of course, content
ourselves with a restricted variation in the Hilbert space. For instance, we
can restrict the variational Hilbert space to wave functions $| \Phi \rangle$
 satisfying $ \langle \Phi | \hat{N} | \Phi \rangle = N_0$. In this case 
 the solution to  the Kamlah second order approximation is given by 
\begin{equation}
\frac{\delta}{\delta \Phi}\langle \hat{H} - h_2 (\Delta\hat{N})^{2}\rangle = 0
\label{cran1}
\end{equation}
with the constraint $ \langle \Phi | \hat{N} | \Phi \rangle = N_0$, where $h_2$
now {\it must be varied during the minimization process}. This solution, however, may not provide 
the deepest $E_{proj}^{(2)}$.
In other words the wave function providing  the energy minimum of the SCK2
method does not necessarily  satisfies
$ \langle \Phi | \hat{N} | \Phi \rangle = N_0$.
 One must be cautious about this apparent freedom on the value 
of  $\langle \hat{N} \rangle$, because an unconstrained variation of
$\langle \hat{N}\rangle$, in general, amounts to a variation of 
$\langle (\Delta \hat{N})^2 \rangle$. Since we have assumed in the derivation
of the Kamlah expansion large  $\langle (\Delta \hat{N})^2 \rangle$,
and there is no way to restrict the variations of $\langle \hat{N}\rangle$ to
values satisfying this condition, it does not seems a good idea to allow free
variation of $\langle \hat{N}\rangle$. In order  not to spoil
the quality of the Kamlah expansion 
we shall keep the condition $\langle \hat{N}\rangle= N_0$ during the variation
of eq.~(\ref{cran1}).
 In refs.~\cite{ZSF.92,DN.93} the same
restriction was made to solve the SCK2 equations. The general expression of
$\delta h_2$ and the specific one for the Gogny force can be found in
ref.~\cite{VAL.96}. From this viewpoint we clearly see that the LN recipe
is an approximation to the SCK2 method, since both minimize eq.~(\ref{cran1})
with the constraint on the number of particles, only in the SCK2 
the coefficient $h_2$ is varied and in the LN is not. 
The wave function that minimizes the LN energy is determined by
\begin{equation}
\delta \langle \hat{H} \rangle - h_2 \; \delta \langle (\Delta\hat{N})^{2}\rangle = 0,
\end{equation}
with the constraint $\langle \hat{N} \rangle = N_0$. Whereas the SCK2 solution
is determined by
\begin{equation}
\delta \langle \hat{H} \rangle - 
 h_2 \; \delta \langle (\Delta\hat{N})^{2}\rangle
- \langle (\Delta\hat{N})^{2}\rangle \; \delta  h_2 = 0,
\end{equation}
with the same constraint on $\hat{N}$. In general, the LN and the SCK2 solutions
are different, only in the special case when $h_2$ is nearly constant, 
i.e. $\delta  h_2 =0$, both solutions do coincide.

 From the fact that the LN
approach is an approximation to the SCK2 one can not necessarily conclude 
that the SCK2 is better in all pairing regimes.
One can say that if both solutions
have large $\langle (\Delta\hat{N})^{2}\rangle $, as to ensure a good convergence
of the Kamlah expansion, then the SCK2 solution is better than the LN one.
However, in the cases where the LN fluctuations are larger than the SCK2
ones, one can not say anything because the quality of the expansion might
be different in both approaches. 

To calculate high spin states
one substitutes $\hat{H}$ by $\hat{H} - \omega \hat{J}_x$ in the variational
equations and impose the additional constraint 
$\langle J_x \rangle = \sqrt{I(I+1)}$.

The derivation above applies for non-density dependent interactions.
 The use of density dependent Hamiltonians in approximate  and exact
projected theories poses
the problem of which density has to be used in the Hamiltonian in the evaluation
of non-diagonal matrix elements of the Hamiltonian, $\langle\Phi| H |\Phi^\prime \rangle$.
In  a recent letter \cite{VER.97} we have proposed  a consistent description
of the  Kamlah (Lipkin-Nogami)  approximation for density dependent 
Hamiltonians. In the
new formalism the projected energy is still given by eq.~(\ref{energy2}) but the
coefficient $h_2$ has now  to be  substituted by an effective parameter
$h_{2}^{eff} $ given by

\begin{eqnarray}
&& h_{2}^{eff} = \frac{(\langle \hat{\cal{H}}  
-\langle  \hat{\cal{H}} \rangle )(\Delta\hat{N})^{2}\rangle
- \langle \hat{\cal{H}} \Delta\hat{N}\rangle\langle(\Delta\hat{N})^{3}\rangle
/\langle(\Delta\hat{N})^{2}\rangle}
{\langle(\Delta\hat{N})^{4}\rangle - \langle(\Delta\hat{N})^{2}\rangle^{2} 
- \langle(\Delta\hat{N})^{3}\rangle^{2}/\langle(\Delta\hat{N})^{2}\rangle},
\end{eqnarray}
with
\begin{equation}
\hat{\mathcal{H}} = \hat{H} +
 \sum_{ij} \langle \frac{\partial\hat{H}}{\partial\rho(\vec{r})}
f_{ij}(\vec{r}) \rangle c^+_i c_j.
\label{heff}
\end{equation}
The quantities $f_{ij}(\vec{r})$ are those appearing in the second
quantization form of the density operator $\hat{\rho}(\vec{r})=\sum_{ij}
f_{ij}(\vec{r}) c^+_i c_j$. The last term in  eq.~(\ref{heff}) is a
consequence of the density dependence of the Hamiltonian and does not
appear in the standard derivation of the Lipkin-Nogami approximation.
It resembles the usual rearrangement term appearing in the HFB theory
with density dependent forces.
It is important to notice that in $h_2$ the full hamiltonian has to be
considered.

\section{Results}
\label{secres}

To solve the LN and the SCK2 equations we have expanded the quasiparticle
operators
of the Hartree-Fock-Bogoliubov transformation in a  triaxial harmonic
oscillator (HO) basis.
  The  HO configuration space was determined by the condition
\begin{equation}
\hbar \omega_x n_x + \hbar \omega_y n_y + \hbar \omega_z n_z \leq \hbar \omega_0 N_0
\end{equation}
where $n_x, n_y$ and $n_z$ are the HO quantum numbers and the frequencies
$\hbar \omega_x, \hbar \omega_y$ and  $\hbar \omega_z$ are determined by
$\omega_x = \omega_y = \omega_0 q^{\frac{1}{3}},\;
 \omega_z = \omega_0 q^{-\frac{2}{3}}$.
The parameter $q$ is strongly connected to the ratio between the nuclear
size along  the $z$- and  the perpendicular direction. In the calculations 
a value of $q=1.5$ 
is used for superdeformed  nuclei and $q=1.3$ for normal deformed ones. 
For $ N_0$ we have taken a  value of  $N_0 =12.5$ for superdeformed
states and $N_0 =11$ for normal deformed ones, 
which provide a basis big enough as to warrant the convergence of the results. 
In our calculations we use the Gogny force with the usual parameterization D1S 
\cite{BGG.84,BGG.91}. 
  As it is usually done, to save CPU time, the following terms of the 
interaction have not been taken into account in the calculations~:
 The Fock term of the Coulomb interaction and the contributions to the 
 pairing field of the spin-orbit, Coulomb and the two-body center of 
mass correction terms.   
 To solve the LN and SCK2 variational equations the Conjugate Gradient Method  
 of ref.~ \cite{ELM.95} has been properly generalized. 

\begin{figure}[h]
\begin{center}
\parbox[c]{9cm}{\includegraphics[width=9cm,angle=0]{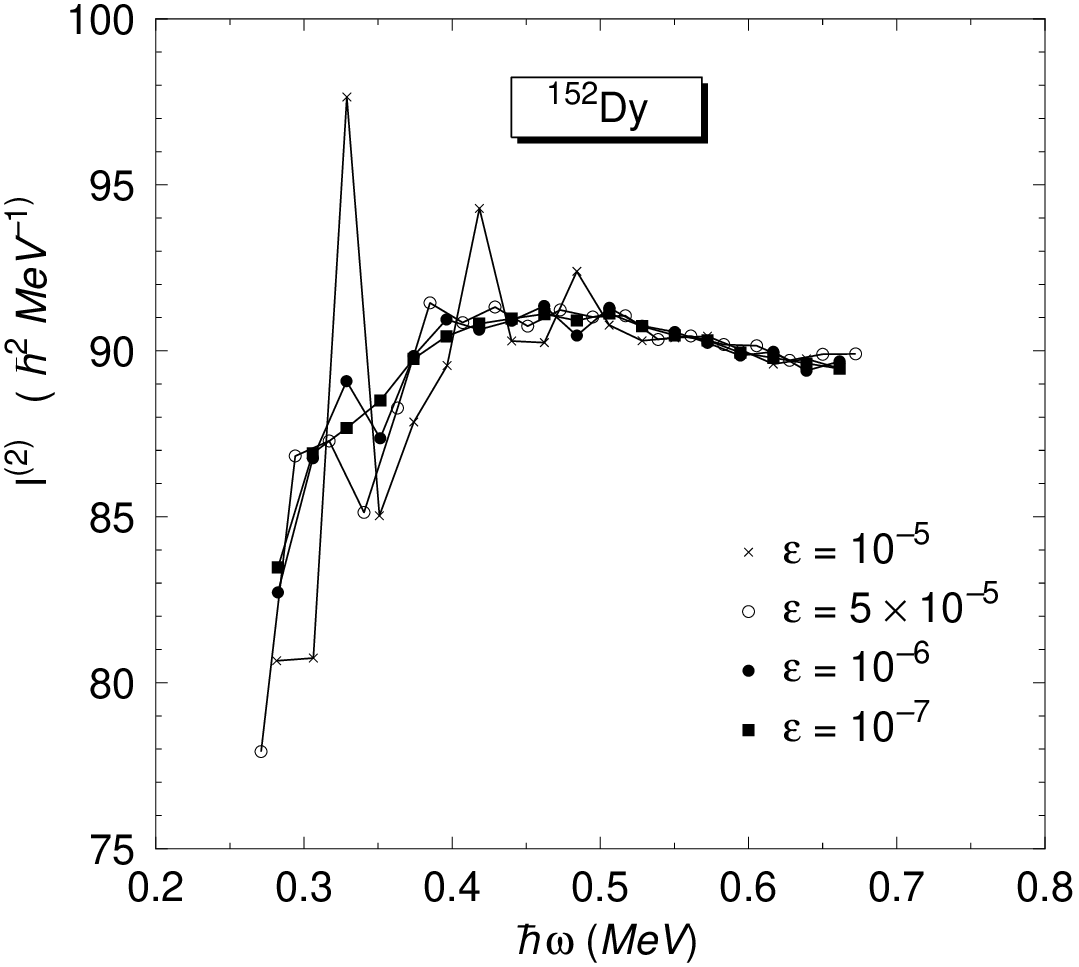}}
\end{center}

\caption{ The dynamical moment of inertia  for different
accuracies in the LN calculations.  $\varepsilon$ is given in MeV. }
\label{convergencia}
\end{figure}

Concerning the convergence of both calculations  the LN  approach converges much 
more slowly than the SCK2 one. This is due to the non-variational character
 of the LN method. The criterion to end the iterative process is that the 
 energy difference
between two consecutive line minimizations  is smaller than a given
parameter $\varepsilon$. In 
selfconsistent calculations, like HF, HFB and SCK2 it is enough to take 
$\varepsilon$ of
the order of $10^{-3}-10^{-4}$ MeV. To illustrate this point we show in 
Fig.~\ref{convergencia} the dynamical
moment of inertia of the superdeformed ground band of the nucleus
$^{152}$Dy for different $\varepsilon$ values. We find that at small angular
momentum, i.e. at the large pairing correlations regime (see below) 
a very high accuracy is required in order to achieve convergence. At high
spins (weak pairing regime), however, the convergence is reached very soon.
As we shall see below this fact might be related with problems of the LN
approach to find the right minimum  in the presence of large pairing
correlations.  
 All calculations reported in this paper in the LN approximation
have been done with  $\varepsilon = 5 \times 10^{-5}$ MeV.
  
\subsection{The Yrast band of $^{164}$Er} 

The first example we would like to show is the  Yrast line 
of the nucleus $^{164}$Er. It is well known that this nucleus presents a 
backbending at around spin $16 \hbar$, due to the alignment of a neutron pair.
This alignment causing a quenching of the neutron pairing correlations at 
high spins. This nucleus is  therefore a very stringent test for the LN prescription and 
the SCK2 approximation.
   
\begin{figure}[h]
\begin{center}
\parbox[c]{14cm}{\includegraphics[width=7cm,angle=270]{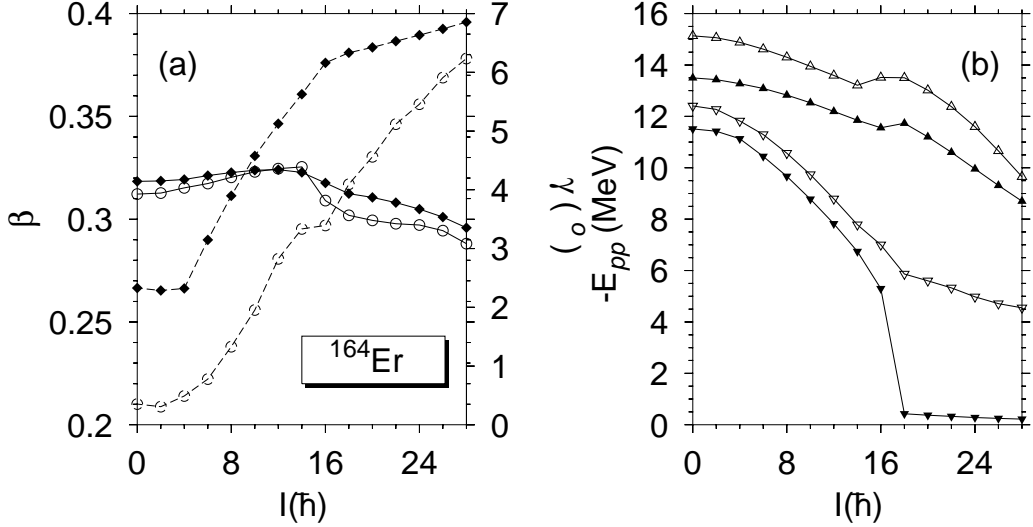}}
\end{center}

\caption{Comparison of LN and SCK2 results versus angular
momentum for $^{164}Er$ along the Yrast band. (a) $\beta$ (full lines, 
left hand side scale)  and $\gamma$ (dashed lines, right hand side scale) 
deformation parameters.
The full (open) symbols correspond to the SCK2 (LN) approximation. (b) Pairing
energies~: full (empty) triangles correspond to protons  in
the SCK2 (LN) method. Inverted full (empty) triangles correspond
to neutrons in the SCK2 (LN) method.  }
\label{betgam}
\end{figure}

 In Fig.~(\ref{betgam}a) we display the $(\beta, \gamma)$ deformation
parameters of the rare earth nucleus $^{164}Er$ along the Yrast line
in the LN and SCK2 methods. In both approximations, up to $I\approx 14 \hbar$,
 we find a slight increase of the $\beta$ deformation parameter caused by the 
Coriolis antipairing effect. Above $I\approx 14 \hbar$ and with increasing
spin values we observe  a decreasing of the $\beta$ deformation
 due to the antistretching effect. 
 Though both approximations provide results close to each other the largest
discrepancies appear for spin values above $I\approx 16 \hbar$.  
 The $\gamma$ deformation increases
with the angular momentum in the LN and the SCK2 methods, the triaxiality
of the SCK2 results being larger than the LN ones.
Fig.~(\ref{betgam}b) displays the particle-particle correlation energy
($E_{pp}=\frac{1}{2} Tr(\Delta \kappa)$) for both methods along 
the Yrast band. 
The proton system behaves very similarly in both approaches,
illustrating very clearly the rotation induced Coriolis antipairing effect, 
 the SCK2 approach providing, though, somewhat smaller values  than the LN one.
The neutron system, however, behaves differently~: while the results for
  $E_{pp}$ in the LN approach decrease
from $-12 $MeV at low spin  to $-6$ MeV at spin value of $20 \hbar$ and 
then change the slope to decrease more gentle, up to $-4$ MeV at spin
$30 \hbar$; the SCK2 values start at $-12 $MeV at low spin  decreasing
to $-6$ MeV at a spin value of $16 \hbar$, at this spin value they experiment a 
sharp drop to an almost unpaired regime.

\begin{figure}[h]
\begin{center}
\parbox[c]{14cm}{\includegraphics[width=7cm,angle=270]{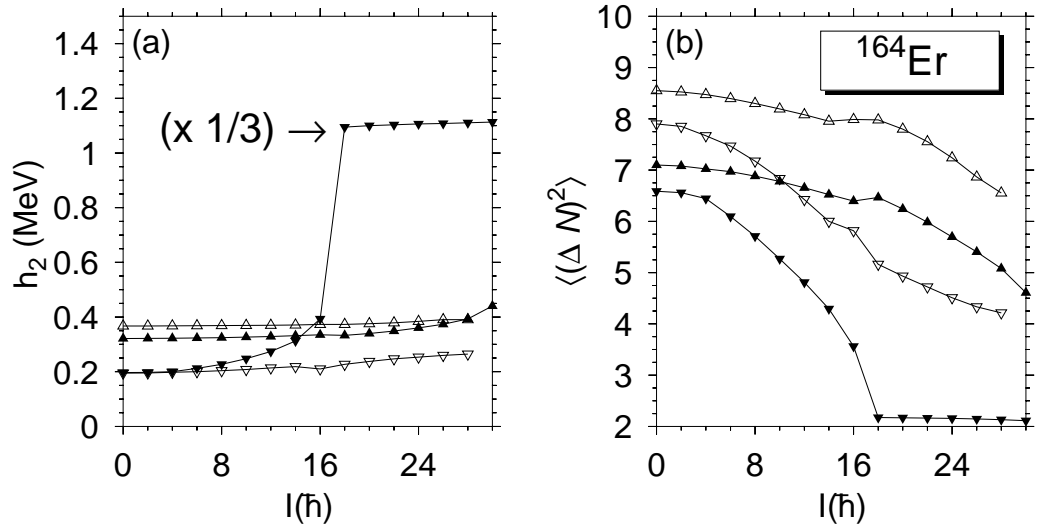}}
\end{center}
\caption{Comparison of LN and SCK2 results versus angular
momentum for the  Yrast band of $^{164}Er$.  (a) $h_2^{eff}$ parameters~:
 Full (empty) triangles correspond to protons  in
the SCK2 (LN) method. Inverted full (empty) triangles correspond
to neutrons in the SCK2 (LN) method. (b) Same
as panel (a), but for the standard deviations $\langle(\Delta\hat{N})^{2}\rangle$.}
\label{h2dn2}
\end{figure}

To understand this behavior and to check the validity of the Kamlah expansion
we  display in Fig.~(\ref{h2dn2}) the values of the $h_{2}^{\tau,eff}$ 
parameter
and the particle number fluctuation in both approaches along the Yrast band.
In the LN approach the values of $h_{2}^{\pi,eff}$ and $h_{2}^{\nu,eff}$
remain small and  constant at low and medium spins and increase slightly at high spin
 values. In the SCK2 approach, however,
the $h_{2}^{\nu,eff}$ parameter is small and constant at low spins, increases smoothly
up to a spin value of $16 \hbar$ and then  very suddenly rises to a large
value, larger than one. On the right hand side of the figure we see how the 
particle number
fluctuations in both approximations decrease more or less smoothly with the
exception of the neutron number fluctuations in SCK2 which decrease rather
sharply from about 6.5 to around 2 at spin value of $18 \hbar$.
Both, the large value of $h_{2}^{\nu,eff}$ and the smallness of
 $\langle (\Delta \hat{N})^2 \rangle$ at high spin values are a 
clear indication of a breakdown of the Kamlah expansion to second order at the
corresponding spin values.
In the LN case we may observe a degradation of the expansion but one still 
remains in the correlated  regime.

\begin{figure}[th]
\begin{center}
\parbox[c]{14cm}{\includegraphics[width=8cm,angle=270]{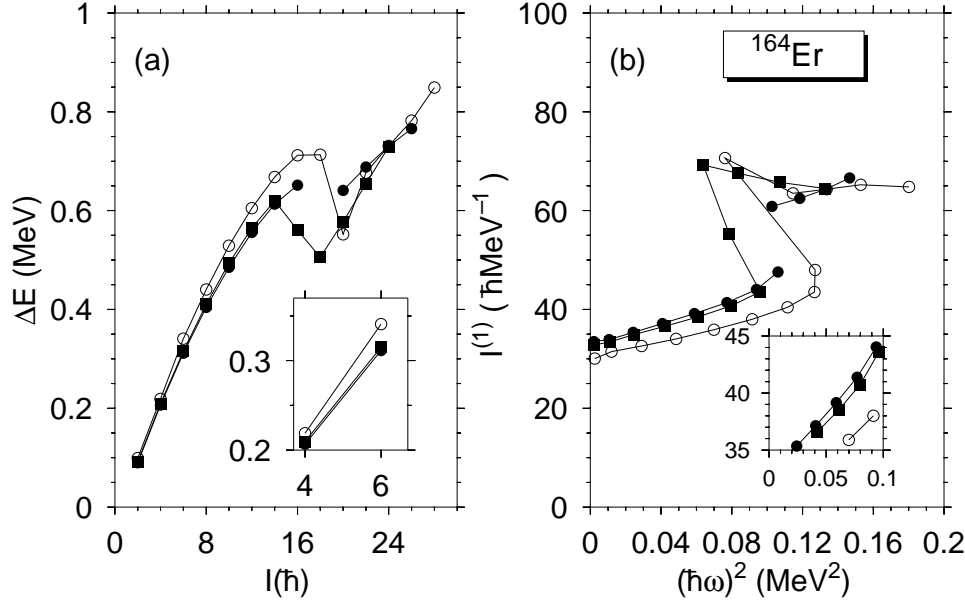}}
\end{center}
\caption{ (a)  Transition energies along the Yrast band of $^{164}$Er
 as a function of the angular momentum. LN (open circles), SCK2 (full circles)
  and experimental data (full squares). 
(b) Comparison of the kinematic moments of inertia  versus the square of 
the angular frequency. Same symbols as (a). }
\label{momer}
\end{figure}

In Fig. \ref{momer}, panel (a), we present the transition energies, 
$\Delta E_I = E(I) - E(I-2) $,
versus the angular momentum, $I$, and in panel (b) the static 
moment of inertia
 $ {\mathcal{I}}^{(1)}=(2I-1)/\Delta E_I $  versus the square
of the angular frequency ($\omega(I)= \Delta E_I/2$), for $^{164}Er$ 
in the LN and the SCK2
approximate methods as well as the experimental data \cite{SWY.80}. 
The LN results are in overall fair agreement with the
experiment. Below the backbending region the pairing correlations are
slightly over-valued. In the backbending region, as expected, the results are
in poorer agreement with the experiment due to the well known fact that
in the cranking approach 
the ground and aligned bands interact at fixed $\omega$ instead of fixed $I$.
Above the backbending region the agreement with the experiment becomes
again good. Concerning the  SCK2 results, for the $I$-values 
 below the backbending, where the  Kamlah expansion is valid, one obtains a 
 better  agreement with the experiment that the LN approach.
 At $I=16 \hbar$, where the Kamlah expansion breaks down,
 we get a huge change in the transition energy (not shown in the picture),
much more pronounced than in the experiment. Beyond this spin value the SCK2 
results are not as good as the LN ones.
In panel (b) of Fig.~\ref{momer}, the backbending plot, ${\cal I}^{(1)}$
versus $ (\hbar \omega )^2 $, is displayed, here we can clearly see that in
the region where the SCK2 method works the agreement with the experiment 
is better than in the LN approach. Outside this region the LN works again
better.  Of course, the Yrast line of the $^{164}$Er nucleus is a very stringent
case because of the  backbending and we expect that the SCK2 approach will work
in other situations.

\subsection{Superdeformed ground band of the nucleus $^{190}$Hg.}

\begin{figure}[th]
\begin{center}
\parbox[c]{18cm}{\includegraphics[width=10cm,angle=270]{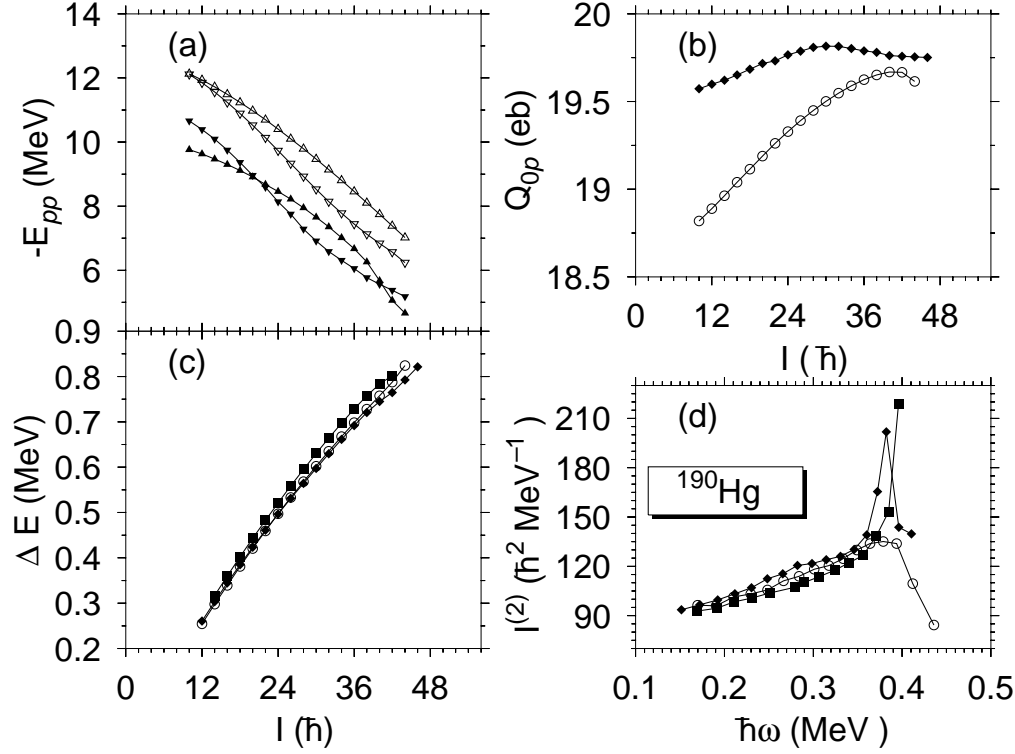}}
\end{center}
\caption{ LN and SCK2 results for the superdeformed ground band of the nucleus
$^{190}Hg$ mercury isotope.
  (a) Pairing energies versus angular momentum~: Full (empty) triangles 
correspond to protons  in
the SCK2 (LN) method. Inverted full (empty) triangles correspond
to neutrons in the SCK2 (LN) method.
 (b) Charge quadrupole moments versus angular momentum~: open circles 
 (full diamonds)  represent the LN (SCK2) results.
(c) Transition energies versus angular momentum~: open circles (full diamonds) 
represent the LN (SCK2) results. The experimental values are depicted 
by full squares.
  (d) Dynamic Moment of inertia $\protect {\mathcal{I}}^{(2)}$ versus the
   angular frequency $\hbar\omega$~: same symbols as in panel (c). }
 \label{190Hg}
\end{figure}

Now we shall discuss the superdeformed ground band of the nucleus $^{190}$Hg as 
an example of  superdeformation in the  $A\simeq 190$ region. 
Calculations in this region in the LN approach have been performed with the
Skyrme force and a density dependent pairing force in \cite{THB.95} and with
the relativistic mean field theory in \cite{Ring190}. 
In panel (a) of Fig.~\ref{190Hg} we display the particle-particle correlation energy
in the LN and SCK2 approaches. As before we obtain less correlations in the 
SCK2 approach than in the LN one, both approaches displaying the typical rotation 
induced Coriolis antipairing effect. For this nucleus, for all spin values,
and in both approximations,  the pair correlations  are large enough 
to warrant the convergence of the Kamlah expansion.

\begin{figure}[th]
\begin{center}
\parbox[c]{14cm}{\includegraphics[width=\textwidth,angle=0]{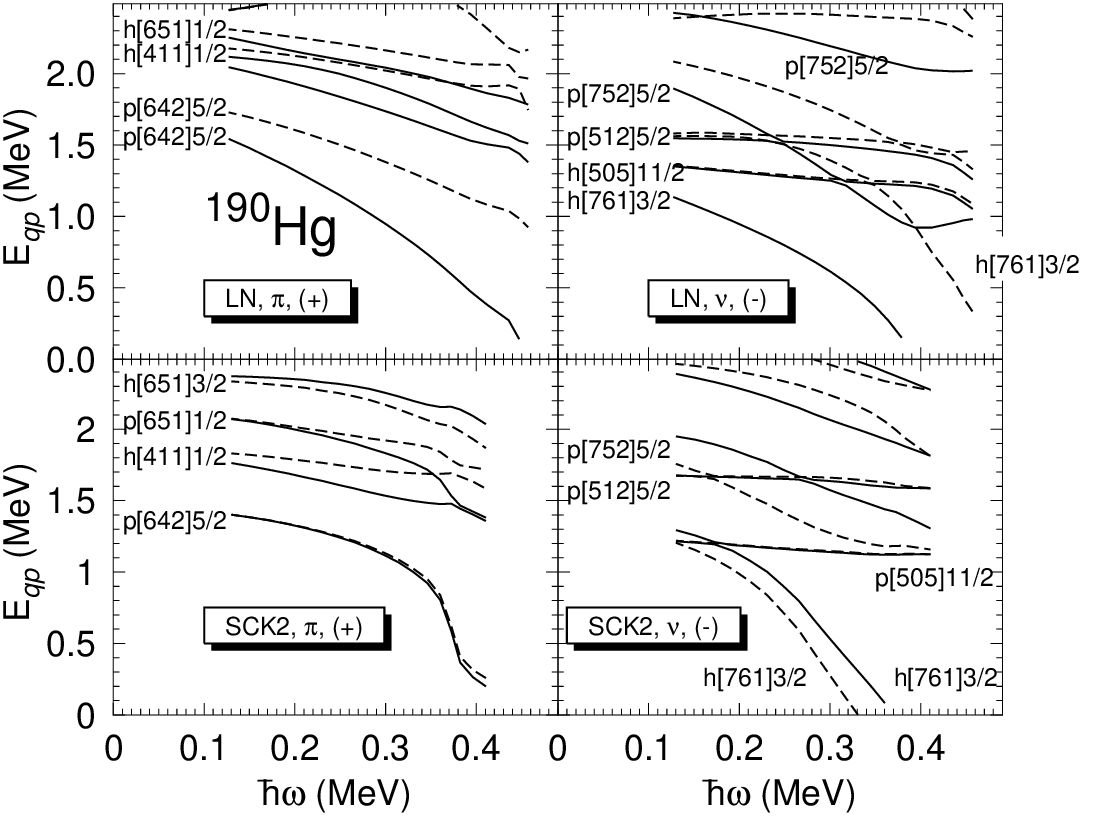}}
\end{center}
\caption{ Quasiparticle energies  versus the
angular frequency $\hbar\omega$ for the SD ground state band of  $^{190}Hg$.
 Only the positive parity proton and negative parity neutron orbitals are
 shown. Orbitals of positive (negative) signature are represented by solid
 (dashed) lines.}
\label{190Hgqp}
\end{figure}

In  panel (b) the charge quadrupole moments are shown. At low spin
values the SCK2 values are somewhat larger than the LN ones, due in part 
to the fact that the latter ones are more pair-correlated, while at higher
spins they get closer.  Both predictions are slightly larger than the experimental
value of $17.7^{+1}_{-1.2}$ eb \cite{AMR.97}. The transition energies
$\Delta E_I$ are shown in panel (c), at low and medium spins both approaches
are very close to each other and in good agreement with the experimental data
\cite{CRO.95}.
At high spins, though the overall agreement with the experimental data
is better in the LN approach, a careful look reveals that the down bending
of the experimental two last points is not provided in this approach. The
SCK2 method, however, describes this feature. This behavior can be more clearly
seen in the plot of the second moment of inertia, 
${\cal I}^{(2)}= 4/(\Delta E_I - \Delta E_{I-2})$,
in panel (d). Here we observe that  the theoretical predictions closely follow
the experimental data up to $\hbar\omega \simeq 0.35$ MeV. From this point
on  the experimental data display an upbending that the LN results do
not follow, the SCK2 results, however, clearly show the same behavior as in the
experiment.  We can understand the reason for this upbending looking at
Fig.~\ref{190Hgqp}, where  the quasiparticle energies versus
the angular frequency for  the relevant channels in the LN and SCK2
approaches are plotted. The main difference between both approaches is that in the LN case
the splitting of the signature partner levels $\nu[761]3/2$ and 
$\pi[642]5/2$, close to the Fermi surface, is larger than in the SCK2 case. 
The signature partner  levels $\nu[761]3/2$ align at 
$\hbar\omega \simeq 0.35 MeV$ in the SCK2 case, in the LN approach there is 
a delay in the alignment of the negative signature level with respect to the 
positive one.
We have checked the parameter $h_2^{eff}$ and the fluctuations  
$\langle (\Delta \hat{N})^2 \rangle$ and the Kamlah expansion does not break
down in this case.

\subsection{Superdeformed ground band of the nucleus $^{152}$Dy}

\begin{figure}[th]
\begin{center}
\parbox[c]{18cm}{\includegraphics[width=10cm,angle=270]{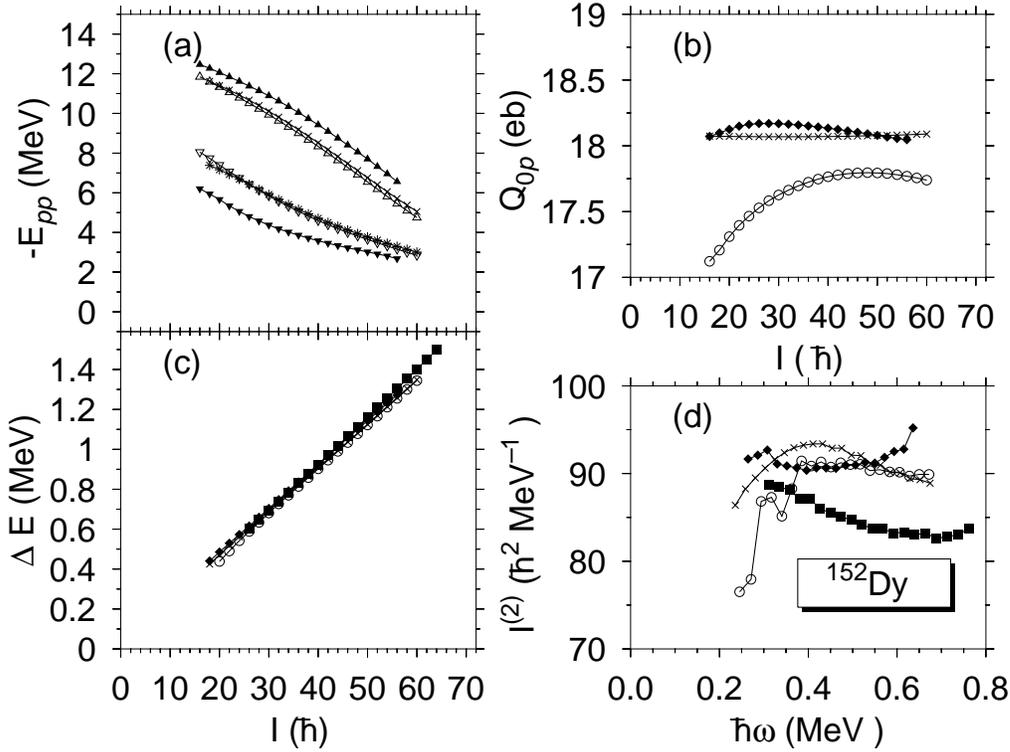}}
\end{center}
\caption{ Same as Fig. 4 but for the nucleus $^{152}$Dy. The additional
symbols are explained in the text.}
\label{fig:kampap4}
\end{figure}

Next we shall turn to discuss an example of a SD nucleus in the $A\sim 150$
region, namely $^{152}$Dy.
Calculations in this region in the LN approach have been performed with the
Skyrme force and a density dependent pairing force in \cite{BFH.96}. 
  As before, we have calculated  the ground superdeformed band
 in the  LN and in the SCK2 approaches.
In Fig.~\ref{fig:kampap4}a we display the particle-particle correlation energy
along this band. The  behavior of $E_{pp}$ with increasing
angular momentum is similar to the ones in the already discussed  nuclei.  We have to notice,
however,  that at high angular momentum the  correlations are small and
that this  nucleus is, so far, the only case, in our calculations, where the 
proton 
pair correlations are larger in the SCK2 approach than in the LN one.
In panel (b) the charge quadrupole moment is depicted, again the largest
difference between both approximations takes place at low spins where
the pair correlations differ at most, at high spins they get closer.
The SCK2 moments are rather constant with spin, as one would expect for a 
superdeformed band,  while the LN ones vary somewhat more.
 The LN results are in good agreement with the experimental data of  
 $17.5 \pm 0.2$ eb
\cite{SKW.96} whereas the SCK2 are slightly higher. In panel (c) the transition 
energies are shown, though both
approaches provide good results as compared with the experimental data
\cite{NJM.97} the SCK2 ones are slightly better. Finally in panel (d) the 
second moment of inertia is plotted versus the angular frequency. 
At medium and high spins the LN approach provides constant values of
 ${\cal I}^{(2)}$ as in the experiment though somewhat larger.
The LN results at low  spins  are  too low  as compared to the
experiment.

At low and medium angular frequencies 
the ${\cal I}^{(2)}$ values in the SCK2 approximation are rather constant with
 spin,  though they are  somewhat larger than the experimental data.
At low spins the SCK2 approximation, however, is in better
agreement with the experiment than the LN approach. At high angular
frequency the  SCK2 results increase too fast.
  Concerning the
convergence of the Kamlah expansion, in the SCK2 at spin values above 
$56 \hbar$ we obtain large values of $h_2$ and very small
$\langle (\Delta \hat{N})^2 \rangle$ indicating that the expansion breaks down.
As a matter of fact, the $h_2$ values of the last  points displayed in the
figure have $h_2$ values somewhat larger than one.
In the LN case the $h_2$ values are always small (around $0.5$) and rather
constant.

 The most striking difference between the LN and the SCK2 results in both
SD nuclei studied  is the behavior of the quadrupole moment at low spins. 
 The SCK2 results look more reasonable because they keep rather constant 
 quadrupole moment with
the spin as one would expect from a superdeformed object. It seems, therefore,
that the  LN values at low spins are less confident. Others, not well 
understood,
facts of the LN approach are the slow convergence  at low spins discussed in 
section \ref{secres} and the behavior of the second moment of inertia at 
low spins in $^{152}$Dy. 
 In order 
to investigate these facts we have performed 
quadrupole moment constrained LN (LNC) calculations. We have calculated
the SD band with the additional constraint that each state of  the band
has a quadrupole moment equal
to a constant value close to the SCK2 at high angular momentum. 
The results are depicted in  Fig.~\ref{fig:kampap4}, all results are represented
by crosses with the exception of the neutron particle-particle correlation
energy  in panel (a) that is represented by asterisks. In this panel we observe
that the results of the  LN constrained calculations for $E_{pp}$ are more or 
less the same as the unconstrained ones. In panel (b) the constant constrained 
quadrupole moment is shown,  this value is closer to the unconstrained LN 
ones at high spins, we expect, therefore,  larger differences between both
approaches at small rather than at large angular momentum.
In panel (c) we see that  the LNC transition energies and the LN ones do 
coincide at high spins,  and that at small spins  the LNC ones are  
closer to the SCK2 and to the experiment than the LN ones. In panel (d), lastly,
 we observe  that the behavior of the second moment of inertia  is more 
reasonable at low spins in the LNC approach  that in the LN one.

\begin{figure}[th]
\begin{center}
\parbox[c]{10cm}{\includegraphics[width=6cm,angle=270]{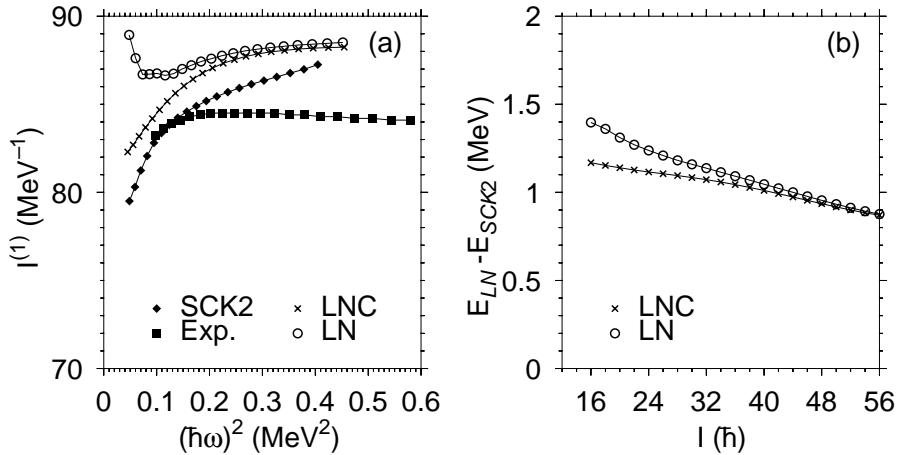}}
\end{center}
\caption{ (a)  Static  moment of inertia versus the square of the angular
 frequency for the SD ground band of $^{152}$Dy. (b) Energy differences between
the LN (LNC) and the SCK2 approaches.}
\label{fig:lnconst}
\end{figure}

   To gain more insight in the different approaches we have represented in 
Fig.~({\ref{fig:lnconst}a) the kinematical moment of inertia 
$\protect {\cal I}^{(1)} $ versus the square of the angular frequency. 
 The experimental results are rather constant at large angular momenta and
they slightly decrease at small angular momentum. The LN results  behave 
properly, though a little too large, at high angular momentum. At small angular
momentum they have an unexpected behavior. The SCK2 results at small angular
momentum behave well and at high $I$  converge to the LN results. 
The LNC calculations on the other hand give the right behavior at small
angular momentum.
In Fig.~({\ref{fig:lnconst}b) we finally present the energy differences between
the LN approaches and the  SCK2. Obviously the SCK2 approach
provides the deepest energy. Concerning the LN and the LNC values, 
it is interesting
to realize that the LNC energies are always deeper or equal to the LN ones.
This is not a situation to wonder because the LN approach is not variational
 and there is no reason, therefore, to expect that by changing some expectation 
 value we must go up, necessarily, in the energy.

This example illustrates clearly that the LN approach, occasionally,
may fail to find the
right minimum because it is not a variational method. 
This failure being, probably, the cause of the slow convergence rate found in
some cases. It is important to notice
that this happens in the strong pairing regime, this behavior was not observed
in earlier investigations \cite{ZSF.92,DN.93} probably because the considered
model was too simplistic.

\section{Conclusions}

In conclusion, for the first time, we have  performed selfconsistent calculations
of approximate particle number projection within the Kamlah
expansion to second order  with density dependent forces (Gogny force). In 
particular, we have reported calculations for the yrast band of the 
well-deformed rare earth nucleus $^{164}Er$, and the superdeformed ground 
 bands of nuclei
$^{190}Hg$ and $^{152}Dy$. In all cases, comparisons have been made
with previous LN calculations and with available experimental data. 
 As a result we find that the Kamlah expansion to second order  may break down 
in  situations of weak pairing
when the minimization equations are solved fully variational (SCK2). In the
Lipkin-Nogami approach, however, this has never happened in the cases analyzed 
so far. 
In the regimes where the expansion is valid, the fully variational approach
provides results closer to the experiment than the non-variational (LN) ones.
 On the other hand, the LN approach has the drawback of the slow convergence 
 in the iteration process and
of not finding the right minimum in some isolated cases in the strong pairing
regime, causing a low convergence rate in the iteration process.

This work was supported in part by DGICyT, Spain under Project  PB97--0023.


\begin{thebibliography}{9}
\bibitem{LIP.60} H. J. Lipkin, Ann. Phys. (N.Y.) {\bf 12}, 425 (1960).
\bibitem{NOG.64} Y. Nogami, Phys. Rev. {\bf B134}, 313 (1964);
                 Y. Nogami and I.J. Zucker,
                 Nucl. Phys. {\bf 60}, 203 (1964).
\bibitem{GN.66} J. F. Goodfellow and Y. Nogami,
                Can. J. Phys. {\bf 44}, 1321 (1966).
\bibitem{QUE.90} P. Quentin, N. Redon, J. Meyer and M. Meyer,
                 Phys. Rev. {\bf C41}, 41 (1990).
\bibitem{ZSF.92} D. C. Zheng, D.W.L. Sprung and H. Flocard,
                 Phys. Rev. {\bf C46}, 1335 (1992).
\bibitem{VER.96} A.Valor, J.L.Egido, L.M. Robledo, Phys. Rev. {\bf C53},
(1996), 172.
\bibitem{VER.99} A. Valor, J.L. Egido, L.M. Robledo,
 Nucl. Phys. {\bf A}, in press.
\bibitem{KAM.68} A. Kamlah, Z. Phys. {\bf 216}, 52 (1968).
\bibitem{DN.93} J. Dobaczewski and W. Nazarewicz,
                 Phys. Rev. {\bf C47}, 2418 (1993).
\bibitem{GOG.75}   D. Gogny, {\em Nuclear Selfconsistent fields}. Eds.
                  G. Ripka and M. Porneuf (North Holland 1975).
\bibitem{RS.80} P. Ring and P. Schuck, {\em The Nuclear Many Body
                  Problem} (1980), Springer--Verlag Edt. Berlin.
\bibitem{VAL.96}  A. Valor, Ph. D. Thesis, Universidad Aut\'onoma de Madrid,
1996, unpublished.
\bibitem{VER.97} A. Valor, J.L. Egido, L.M. Robledo,
Phys. Lett. {\bf B392 }, (1997), 249-254 . 
\bibitem{BGG.84}   J.F. Berger, M. Girod and D. Gogny,
                  Nucl. Phys. {\bf A428}, 23c (1984).
\bibitem{BGG.91} J.F. Berger, M. Girod and D. Gogny,
Comp. Phys. Comm. {\bf 63}(1991) 365-374
\bibitem{ELM.95} J.L. Egido, J. Lessing, V. Martin and L.M. Robledo,
Nucl. Phys. {\bf A594} (1995)70-86 
\bibitem{SWY.80}  S.W. Yates et al.\, Phys. Rev. {\bf C21}, 2366 (1980).
\bibitem{THB.95} J. Terasaki, P.-H. Heenen, P. Bonche, J. Dobaczewski and H. Flocard,
                 Nucl. Phys. {\bf A593} (1995) 1-20 
\bibitem{Ring190} A.V. Afanasjev, J. K\"onig, P. Ring,
   Phys. Rev. {\bf C60}, in press.
\bibitem{AMR.97} H. Amro et al. Phys. Lett {\bf B413} (1997) 15
\bibitem{CRO.95}  B. Crowell et al., Phys. Rev {\bf C51}(1995)  R1599 
\bibitem{BFH.96} P. Bonche, H. Flocard and P.-H. Heenen,
                 Nucl. Phys. {\bf A598} (1996) 169 
\bibitem{SKW.96} H. Savajols et al., Phys. Rev. Lett. {\bf 76} (1996) 4480
\bibitem{NJM.97}  D. Nisius at al., Phys. Lett. {\bf 392B} (1997) 18 

\end{thebibliography}
\end{document}